\documentclass[epj]{webofc}
\usepackage[varg]{txfonts}   
\woctitle{MESON2018 - the 15$^\textrm{th}$ International Workshop on Meson Physics}
\newcommand{\be}{\begin{equation}}
\newcommand{\ee}{\end{equation}}
\begin{document}
\selectlanguage{english}
\title{Near threshold kaon-kaon interaction in the reactions 
$e^+ e^- \to K^+ K^- \gamma$ and
$e^+ e^- \to K^0 \bar{K^0}\gamma$}

\author{L. Le\'sniak \inst{1}\fnsep\thanks{\email{leonard.lesniak@ifj.edu.pl}} \and
        F. Sobczuk \inst{1} \and
        M. Silarski\inst{1} \and
        F. Morawski\inst{1}
}

\institute{Institute of Physics, Jagiellonian University, Cracow, Poland}
\abstract{
Strong interactions between pairs of the $K^+ K^-$ and $K^0 \bar{K^0}$ mesons can be studied in radiative decays of $\phi(1020)$ mesons. We present a theoretical model of the reactions $e^+ e^- \to \phi \to K^+ K^- \gamma$ and $e^+ e^- \to \phi \to K^0 \bar{K^0}\gamma$.
The $K^+ K^-$ and $K^0 \bar{K^0}$ effective mass dependence of the differential cross sections is derived. The total cross sections and the branching fractions for the two radiative $\phi$ decays are calculated.  
}
\maketitle
\section{Description of the theoretical model}
\label{intro}
The kaon-kaon strong interaction near threshold is largely unknown.
Also the parameters of the scalar resonances $f_0(980)$ and $a_0(980)$
are still imprecise.
The $\phi(1020)$ meson decays into $\pi^+\pi^- \gamma$, $\pi^0\pi^0 \gamma$ and $\pi^0\eta \gamma$ have been measured, for the $\phi$ transition into  $K^0 \bar{K^0} \gamma $ only the upper limit of the
branching fraction has been obtained in Ref.~\cite{KLOE0} but there are no data for the $\phi \to K^+ K^- \gamma$ process.
 
In this paper we outline a general theoretical model of the $e^+e^-$ reactions leading to final states with two pseudoscalar mesons
and a photon. 
At the beginning we derive the amplitude $A(m)$ for the $e^+ e^- \to K^+ K^- \gamma$ process.  
It is a sum of the four amplitudes corresponding to diagrams (a), (b), (c) and (d) in Fig.~\ref{fig-1}:
\begin{figure}[ht]
\centering
\includegraphics[width=3.1cm]{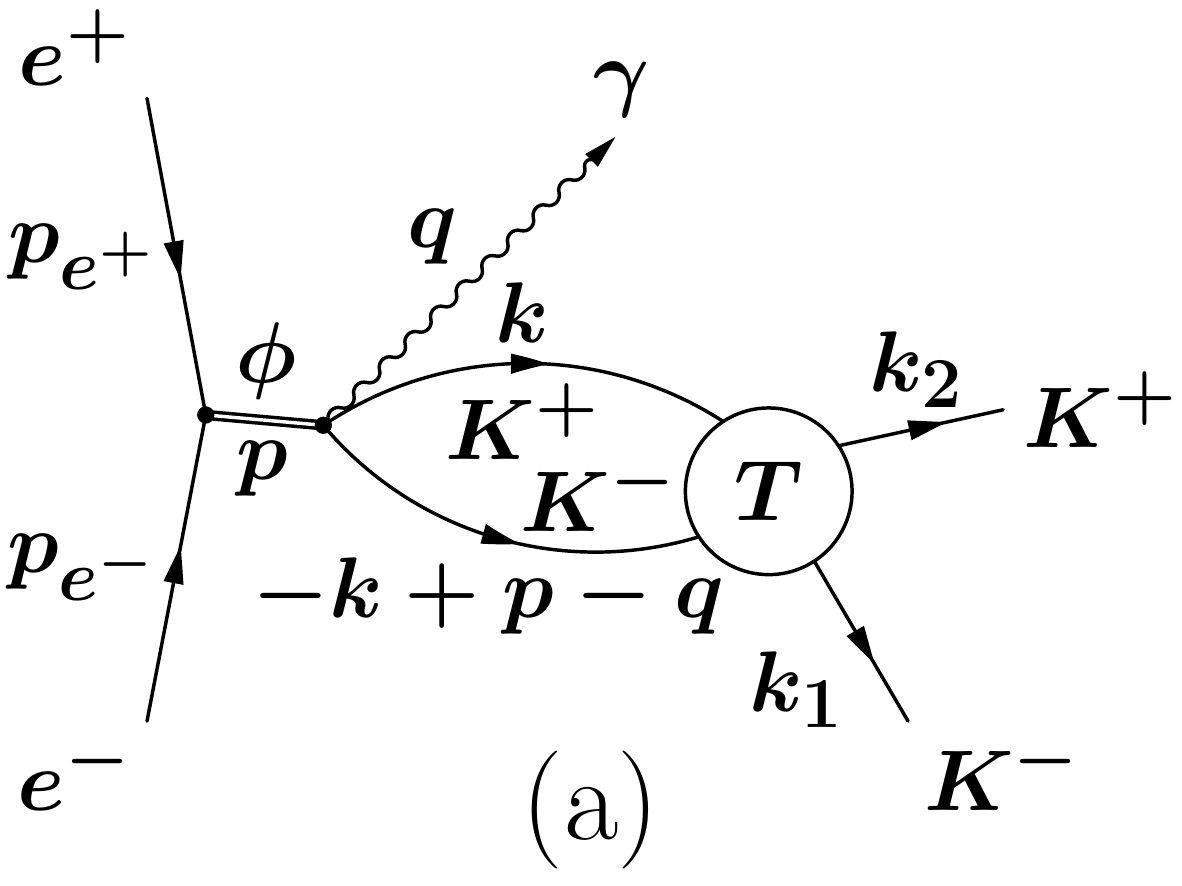}~
\includegraphics[width=3.1cm]{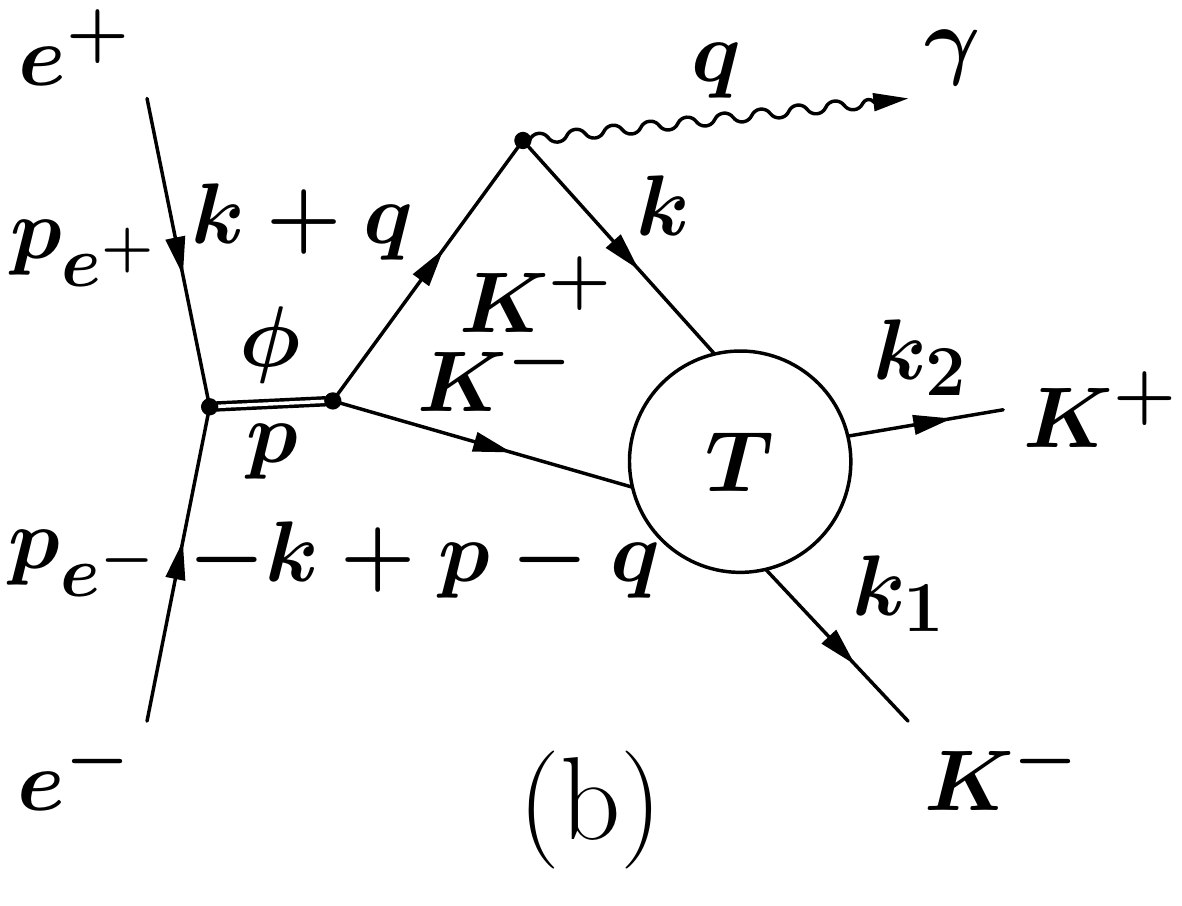}~
\includegraphics[width=3.1cm]{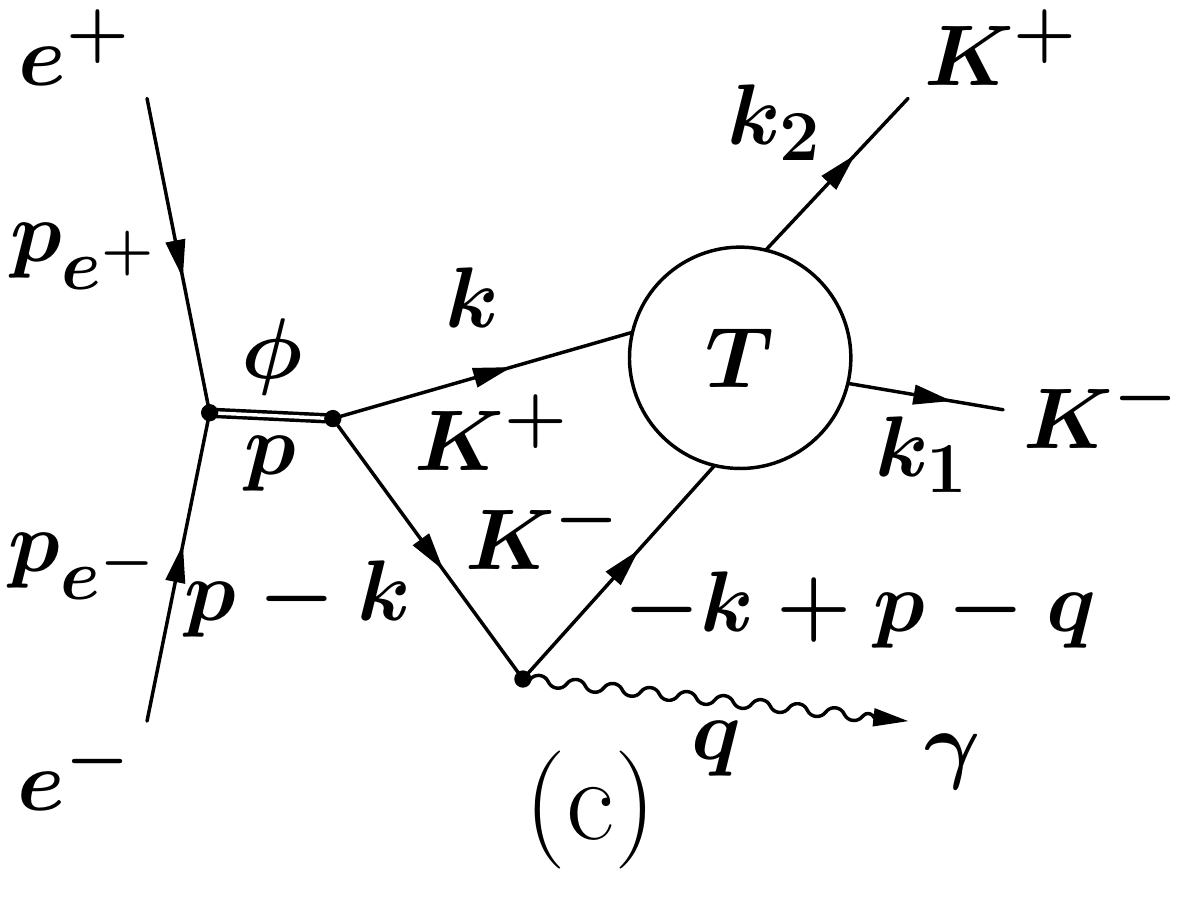}~
\includegraphics[width=3.1cm]{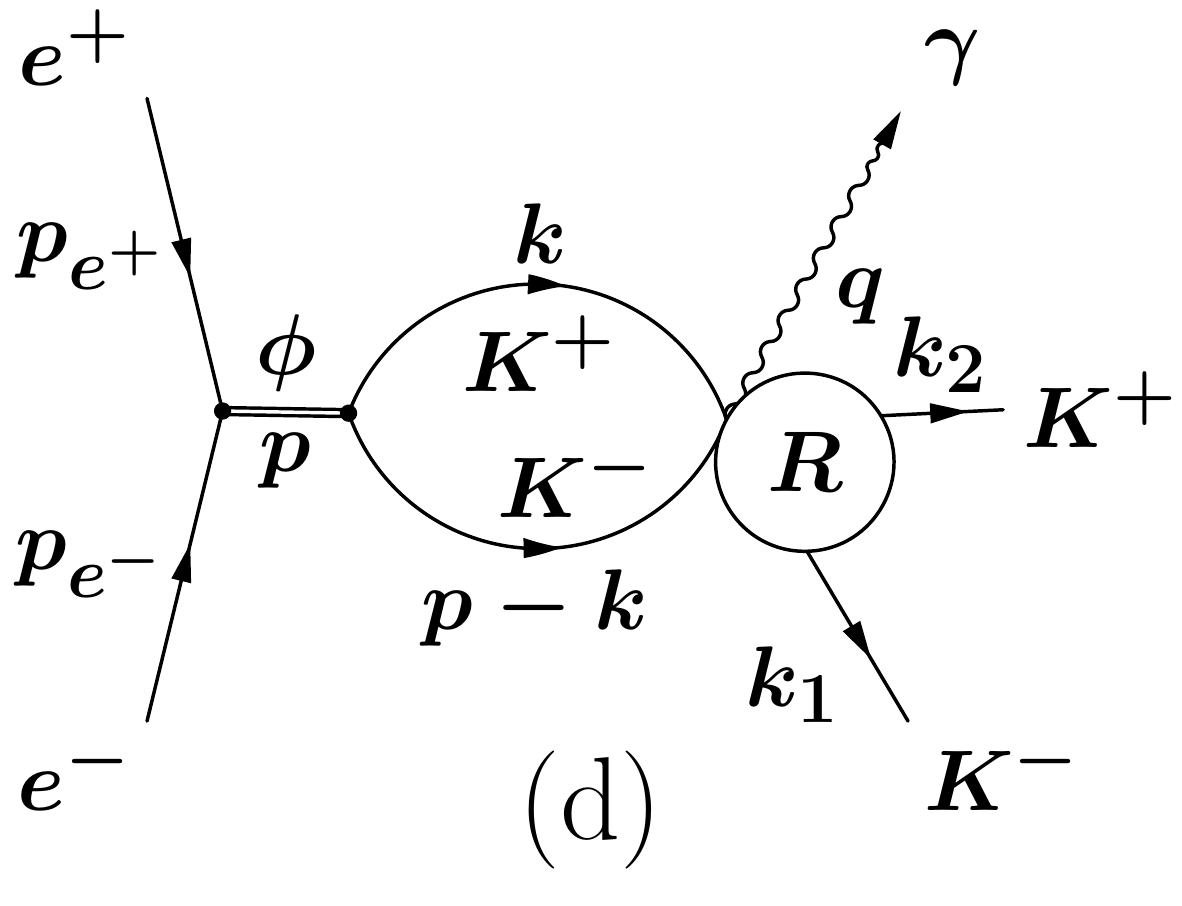}
\caption{Diagrams for the reaction $e^+ e^- \rightarrow K^+ K^- \gamma$ with final-state $K^+ K^-$ interaction.
The $K^+K^-$ elastic 
 amplitude is labelled by $T$ and $R$ denotes the difference of the $K^+K^-$ amplitudes $T$ in Eq.~(\ref{Ad}).
}
\label{fig-1}       
\end{figure}
\be
\label{Aa}
A_a=2 i \int \frac{d^4 k}{(2 \pi)^4} \frac{J_{\nu} \epsilon^{\nu *} T(k)}{D(k) D(-k+p-q)},
\end{equation}
\be
\label{Ab}
A_b=-4 i \int \frac{d^4 k}{(2 \pi)^4} \frac{J_{\mu} \epsilon^{\nu *} k_{\nu} (k_{\mu}+q_{\mu})T(k)}{D(k+q) D(k) D(-k+p-q)},
\end{equation}
\be
\label{Ac}
A_c=-4 i \int \frac{d^4 k}{(2 \pi)^4} \frac{J_{\mu} \epsilon^{\nu *}~ (k_{\nu}-p_{\nu}) k_{\mu} T(k)}{D(p-k) D(k) D(-k+p-q)},
\end{equation}
\begin{equation}
\label{Ad}
A_d=-2 i \int \frac{d^4 k}{(2 \pi)^4} \frac{J \cdot k~ \epsilon^* \cdot \tilde{k}}{D(k) D(p-k)}
\frac{[T(k-q)-T(k)]}{q\cdot \tilde{k}}.
\end{equation}
One can show that the amplitude $A(m)=A_a+A_b+A_c+A_d$ is gauge invariant.
In Eqs.~(\ref{Aa}-\ref{Ad}) $D(k)=k^2-m_{\rm K}^2 + i \delta$, $\delta \rightarrow +0$, is the inverse of the kaon propagator, $m_{\rm K}$ is the charged kaon mass,
the four-vector $\tilde{k}=(0,\hat{\mathbf{k}})$ with the unit three-vector $\hat{\mathbf{k}}=\mathbf{k}/|\mathbf{k}|$. 
In the above expressions $q$ is the photon four-momentum,  $p=p_{e^+}+p_{e^-}$ is the $\phi$ meson four-momentum, $\epsilon^{\nu}$ is the photon polarization four-vector and $J_{\mu}$ is defined as
\be
\label{J}
J_{\mu}=\frac{e^3}{s} F_K(s) \bar{v}(p_{e^+}) \gamma_{\mu} u(p_{e^-}),
\end{equation}
where $e$ is the electron charge, $s=(p^2$ is the Mandelstam variable, $v$ and $u$ are the $e^+$ and $e^-$ bispinors,
respectively, $\gamma_{\mu}$ are the Dirac matrices and
$F_K(s)$ is the kaon electromagnetic form factor.  
The $K^+ K^-$ elastic scattering amplitude is given by
\be 
\label{Tk}
T(k)=\langle K^-(k_1) K^+(k_2)|\tilde{T}(m)|K^-(-k+p-q) K^+(k)\rangle,
\end{equation}
where $m^2=(k_1+k_2)^2$ is the square of the $K^+ K^-$ effective mass and $\tilde{T}(m)$ is the $K \bar K$ scattering operator.
The on-shell $K^+ K^-$ amplitude can be expressed as $T_{K^+ K^-}(m)=\langle K^-(k_1) K^+(k_2)|\tilde{T}(m)|K^-(k_1) K^+(k_2)\rangle$.
The four-momenta of kaons in the $K^+ K^-$ center-of-mass frame are:
$k_1=(m/2,-\mathbf{k_f})$ and $k_2=(m/2,\mathbf{k_f})$, where $k_f=\sqrt{m^2/4-m_{\rm K}^2}$ is the kaon momentum in the final-state. 
We can assume that $T(k)$ is related to $T_{K^+ K^-}(m)$ as follows:
\be
\label{Ton}
T(k)\approx g(k) T_{K^+ K^-}(m),
\end{equation}
where $g(k)$, as a real function of the modulus of the kaon three-momentum $k\equiv|\mathbf{k}|$, takes into account the off-shell character of $T(k)$.
From Eqs.~(\ref{Tk}-\ref{Ton}) one infers that $g(k_f)=1$.

Under a dominance of the pole at $m=2 E_k\equiv 2 \sqrt{\mathbf{k}^2+m_{\rm K}^2}$,
the amplitude $A(m)$ can be written in the following form:
\be
\label{Atot2}
A(m)= ~\vec{J} \cdot \vec{\epsilon}~^*~T_{K^+ K^-}(m)~[I(m)-I(m_{\phi})],
\ee
where the integral $I(m)$ reads
\be
\label{I}
I(m)= -2
~\int \frac{d^3 k}{(2 \pi)^3}
\frac{g(k)}{2E_km(m-2E_k)}
\left[1-2~\frac{|\mathbf{k}|^2-(\mathbf{k\cdot \hat{q})^2}}{2 p_0 E_k-
m_{\phi}^2 +2\mathbf{k\cdot q)}}\right].
\ee
In Eq.~(\ref{I}) $p_0=m+\omega$, where $\omega$ is the photon energy in the $K^+ K^-$ center-of-mass frame, $m_{\phi}$ is the $\phi$ meson mass and $\hat{\mathbf{q}}=\mathbf{q}/|\mathbf{q}|$ is the unit vector indicating the photon direction.

In the model of Close, Isgur and Kumano~\cite{Close} the momentum distribution of the interacting kaons has been expressed by the function
\begin{equation}
\label{fik}
\phi(k)=\frac{\mu^4}{(k^2+\mu^2)^2},
\end{equation}
with the parameter $\mu=141$ MeV.
This function is normalized to unity at $k=0$, however the function $g(k)$ in Eq.~(\ref{Ton}) has to be normalized to 1 at $k=k_f$, so the function $g(k)$ corresponding to $\phi(k)$ should be defined as
\begin{equation}
\label{giek}
g(k)=\frac{(k_f^2+\mu^2)^2}{(k^2+\mu^2)^2}.
\end{equation}
In Ref.~\cite{Close} kaons are treated as extended objects forming a quasi-bound state.
If the $K^+K^-$ system is point-like, like in Refs.~\cite{Ivan} or~\cite{Aczasow2001}, then the function $g(k)\equiv 1$.
Let us note that both models can be treated as special cases of our approach.

Separable kaon-kaon potentials can be used to calculate the kaon-kaon amplitudes needed in practical application of the present model. Then the function g(k) takes the following form:
\be 
\label{gsep}
g(k)=\frac{k_f^2+\beta^2}{k^2+\beta^2},
\ee
where $\beta$ is a range parameter. 
In Ref.~\cite{KLL} for the isospin zero $K \bar{K}$ amplitude the value
of $\beta$ close to 1.5 GeV has been obtained.
In order to get an integral convergence at large $k$ in Eq.~(\ref{I})
we use an additional cut-off parameter $k_{\rm{max}}$=1 GeV.

\section{Numerical results}
\label{sec-2}

\begin{figure}[ht]
\centering
\includegraphics[width=5.8cm]{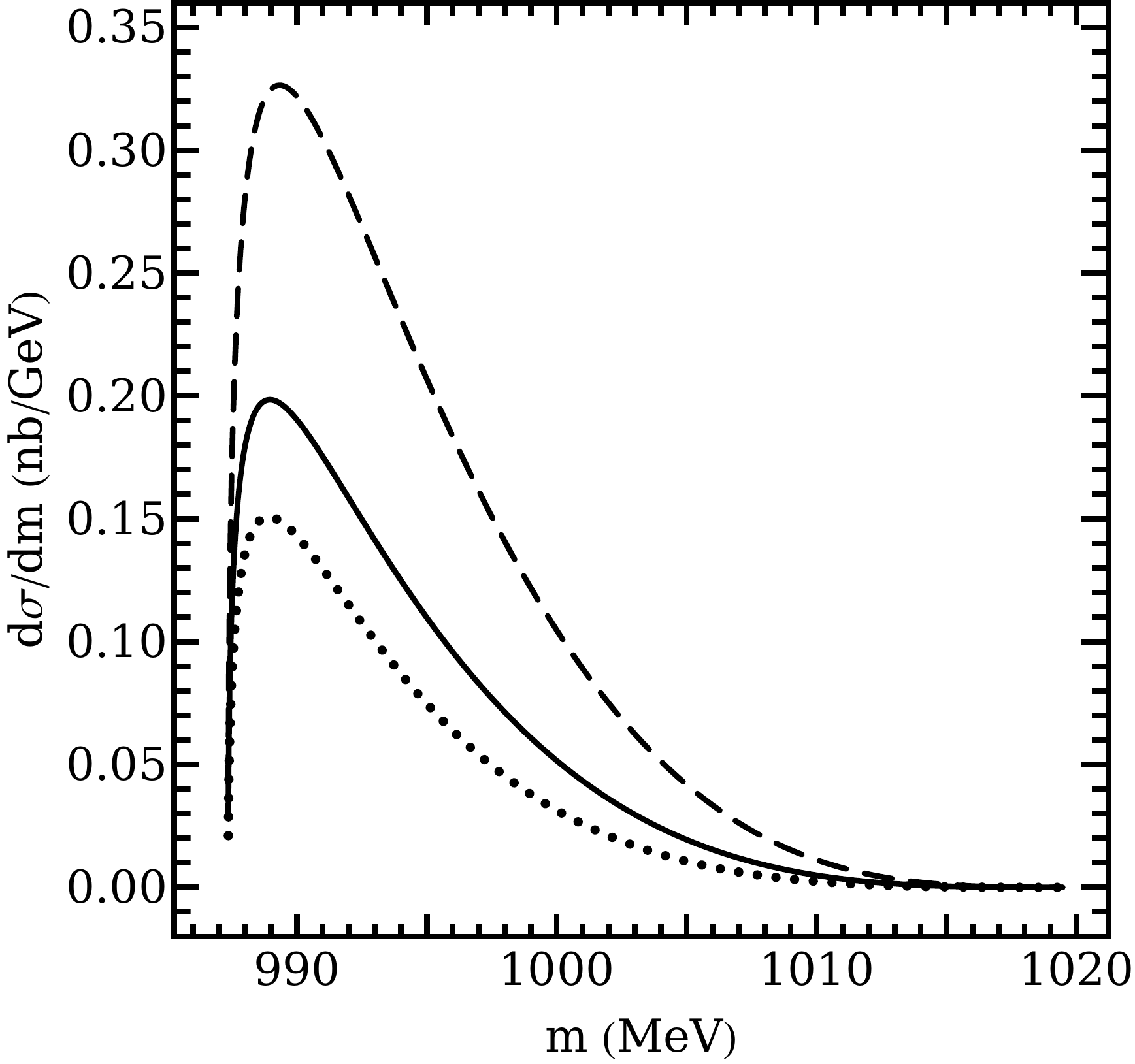}~~~
\includegraphics[width=5.8cm]{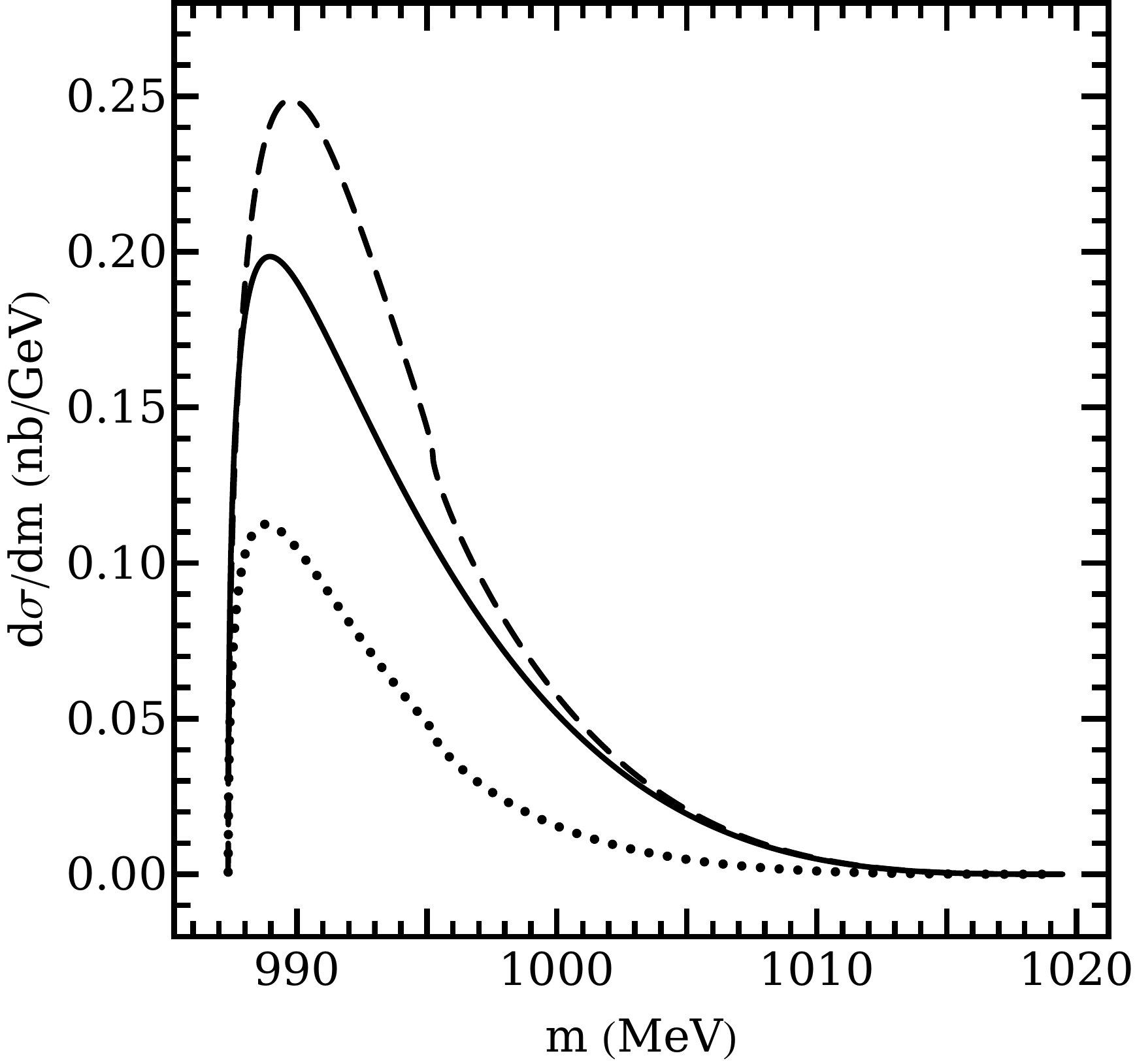}
\caption{Dependence of the differential cross-section for the reaction $e^+ e^- \to K^+ K^- \gamma$ on the $K^+K^-$ effective mass $m$.
 Left panel: the solid line corresponds to the case of the function $g(k)$ (Eq.~\ref{gsep}) with the parameter $\beta \approx 1.5$ GeV and the cut-off $k_{\rm{max}}=1$~GeV, the dotted line - to $g(k)\equiv \phi(k)$ given by Eq.~(\ref{fik})  and the dashed curve - to $g(k)$ from Eq.~(\ref{giek}) with $\mu=141$ MeV;
right panel: the dashed line is calculated for the no-structure model (Ref.~\cite{NS}), the dotted one for the kaon-loop model 
of Ref.~\cite{Aczasow2001} with parameters obtained in Ref.~\cite{KLOE} and the solid line is the same as in the left panel but with a different vertical scale.}
\label{fig-2}       
\end{figure}

In Fig.~\ref{fig-2} we show the $K^+K^-$ effective mass distributions at the $e^+e^-$ energy equal to $m_{\phi}$. On the left panel one observes some dependence on the form of the function $g(k)$. A common feature is a presence of the maximum of the differential cross section situated only a few MeV above the $K^+K^-$ threshold. On the right panel we see a comparison of our model (solid line) with two other models, named the "no-structure" model~\cite{NS} and the "kaon-loop" model, and described
in Refs.~\cite{Ivan} and~\cite{Aczasow2001}. The parameters of the latter models have been taken by us from experimental analysis of the data of the reaction $\phi \to \pi^+\pi^- \gamma$~\cite{KLOE} and then used in calculation of the results shown as dashed and dotted lines.

\begin{figure}[ht]
\centering
\includegraphics[width=5.5cm]{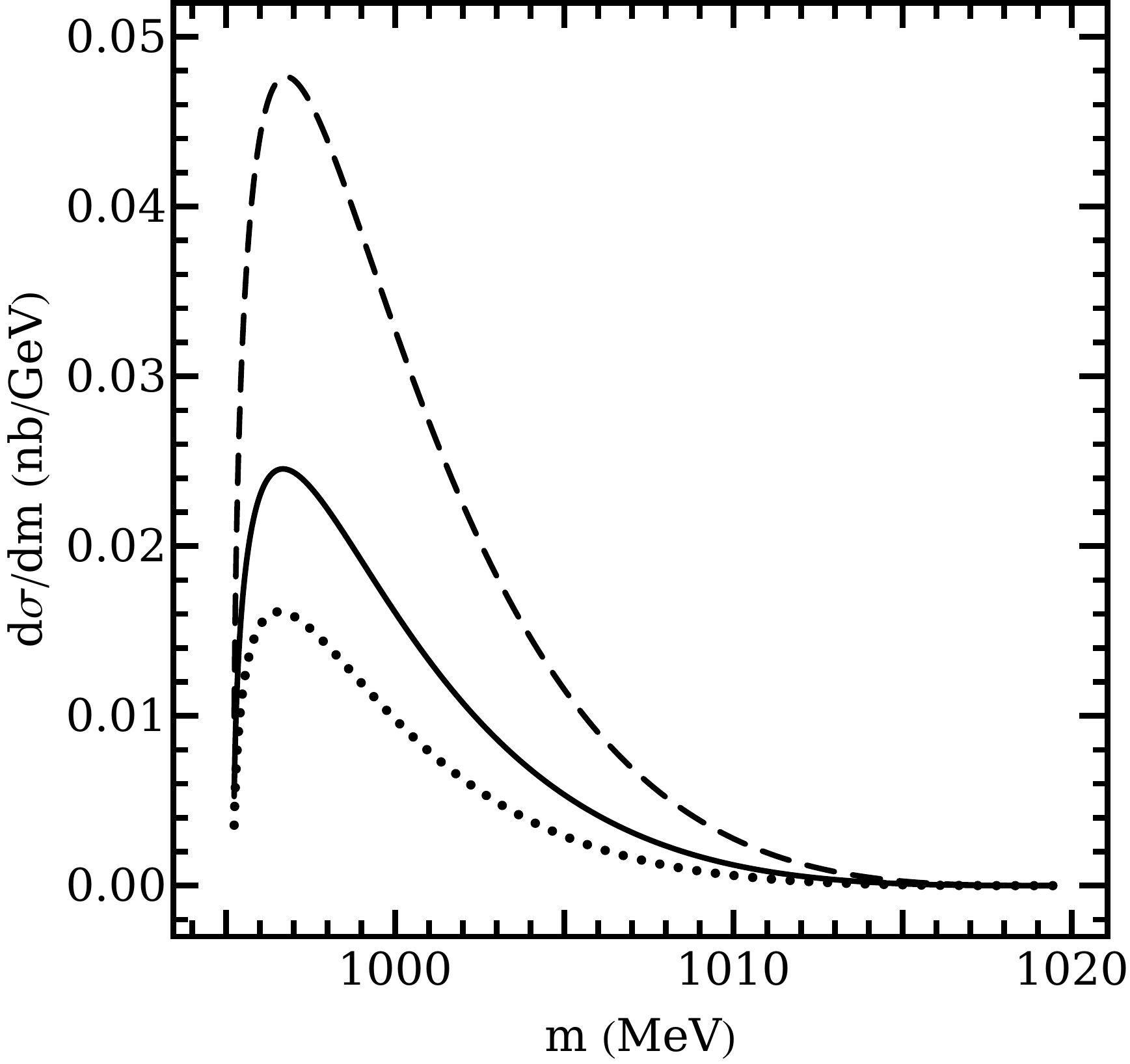}~~~
\includegraphics[width=5.3cm]{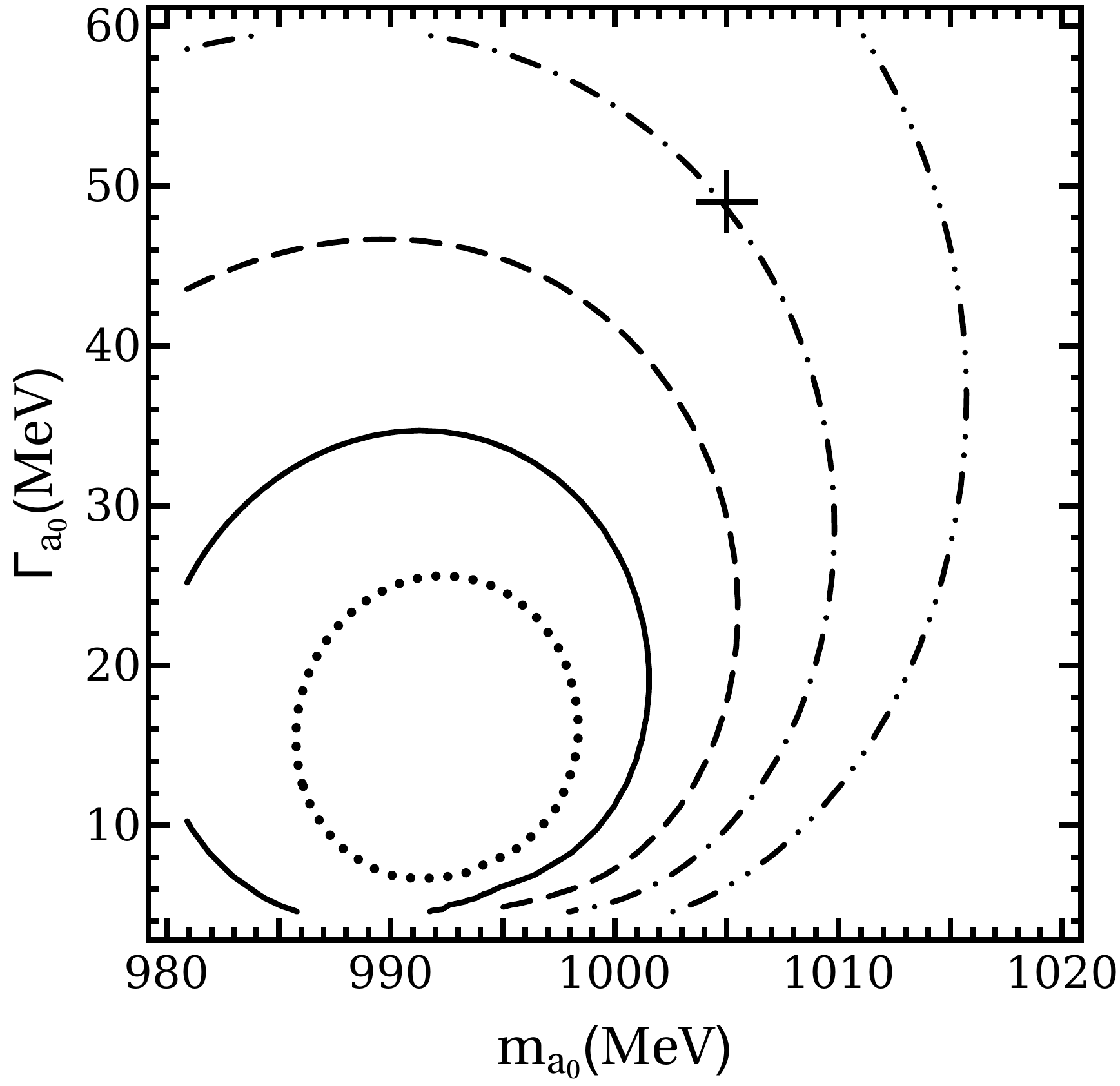}
\caption{Left panel: differential cross-section for the reaction $e^+ e^- \to K^0 \bar{K^0} \gamma$ as a function of the $K^0 \bar{K}^0$ effective mass.
The curves are labelled as in the left panel of Fig.~\ref{fig-2}.
Right panel:
contours of the branching fraction $Br$ for the decay $\phi \to K^0 \bar{K^0} \gamma$ in the complex plane of the $a_0(980)$ pole position:
$m_{a_0(980)}$ is the resonance mass and $\Gamma_ {a_0(980)}$ is its width.
The solid curve corresponds to the KLOE upper limit $Br=1.9 \cdot 10^{-8}$,
the dotted one to $Br=1.0 \cdot 10^{-8}$, the dashed curve to 
$Br=3.0 \cdot 10^{-8}$, the dashed-dotted one to $Br=4.0 \cdot 10^{-8}$ , and the dashed- double dotted one to
$Br=5.0 \cdot 10^{-8}$. The cross indicates the $a_0(980)$ resonance position on sheet $(-+)$ found in Ref.~\cite{Furman}.}
\label{fig-3}       
\end{figure}

The $K^0 \bar{K}^0$ differential cross sections are presented in the left panel of Fig.~\ref{fig-3}. These cross sections are considerably lower than the
$K^+K^-$ cross sections seen in Fig.~\ref{fig-2}. This is due to a smaller $K^0 \bar{K}^0$ phase space and to smaller absolute values of the transition amplitude $T(K^+K^- \to K^0 \bar{K}^0)$ which replaces in this case the elastic $K^+K^-$ amplitude in Eq.~(\ref{Atot2}).

By integration of the $K^+K^-$ and $K^0 \bar{K}^0$ effective mass distributions, shown as solid lines in left panels of Figs.~\ref{fig-2} and ~\ref{fig-3}, one can calculate the total cross sections which are equal to 1.85 pb and 0.17 pb, respectively. The corresponding branching fractions are 4.5$\cdot 10^{-7}$ and 4.0$\cdot 10^{-8}$. 
In the right panel of Fig.~\ref{fig-3} we have plotted the contours of the branching fraction for the $\phi \to K^0 \bar{K}^0 \gamma$ decay
as a function of the $a_0(980)$ resonance position. 
We see that it is possible to generate lower values of the
branching fraction by a moderate change of not well known resonance mass and width.


\section{Conclusions}
\label{concl}
The above theoretical results for the reactions with charged and neutral kaon pairs indicate that the measurements of the $e^+ e^- \to K^+ K^- \gamma$ process
can provide a valuable information about the pole positions of the $a_0(980)$ and $f_0(980)$ resonances.
A coupled channel analysis of the radiative $\phi$ transitions into different
pairs of mesons in the final state is possible after a relevant generalization of the present model. 

\begin{acknowledgement}
This work has been supported by the Polish National Science Centre (grant no 2013/11/B/ST2/04245).
\end{acknowledgement}

\end{document}